\documentclass[a4paper]{scrartcl}
\usepackage[T1]{fontenc}
\usepackage{lmodern}
\usepackage{microtype}
\usepackage[utf8]{inputenc}
\usepackage[pdfborder={0 0 0}]{hyperref}
\usepackage{amsmath,amsthm,amsfonts,mathrsfs}
\usepackage{enumerate,authblk}

\title{Thermal equilibrium states for quantum fields on non-commutative spacetimes}
\author{Gandalf Lechner\thanks{Institut für Theoretische Physik der Universit{\"a}t Leipzig, Gandalf.Lechner@uni-leipzig.de}\; and Jan Schlemmer\thanks{Mathematisches Institut der WWU M\"unster, Jan.Schlemmer@math.uni-muenster.de}}

\newcommand{\N}{\mathbb{N}}
\newcommand{\V}[1]{\mathbf{#1}}
\newcommand{\cV}{\mathcal{V}}
\newcommand{\cT}{\mathcal{T}}
\newcommand{\cP}{\mathcal{P}}

\newcommand{\cD}{\mathcal{D}}

\newcommand{\Schw}{\mathscr{S}}

\usepackage{amsthm}

\newtheorem{Thm}{Theorem}[section]
\newtheorem{Prop}[Thm]{Proposition}
\newtheorem{Lem}[Thm]{Lemma}
\newtheorem{Defn}[Thm]{Definition}

\numberwithin{equation}{section}
\usepackage{oldgerm}
\usepackage{color}
\usepackage{bbm}
\usepackage{amssymb}
\usepackage{epsfig}
\usepackage{amsmath, mathrsfs,psfrag}

\renewcommand{\mathbb}[1]{\mathbbm{#1}}

\newcommand{\Cl}{\mathbbm{C}}
\newcommand{\Rl}{\mathbb{R}}
\newcommand{\Nl}{\mathbb{N}}

\definecolor{lightgray}{rgb}{0.8,0.8,0.8}

\newcommand{\Om}{\Omega}
\newcommand{\om}{\omega}

\newcommand{\te}{\theta}

\newcommand{\la}{\lambda}

\newcommand{\La}{\Lambda}

\newcommand{\J}{\mathcal{J}}

\newcommand{\T}{\mathcal{T}}

\newcommand{\Hil}{\mathcal{H}}
\newcommand{\VV}{\mathcal{V}}

\newcommand{\Pol}{\mathcal{P}}

\newcommand{\Ss}{\mathscr{S}}   

\newcommand{\Cti}{\tilde{C}}

\newcommand{\phiti}{\tilde{\phi}}

\newcommand{\fti}{\tilde{f}}

\newcommand{\gti}{\tilde{g}}

\newcommand{\LGpo}{\mathcal{L}_+^\uparrow}

\newcommand{\LG}{\mathcal{L}}

\newcommand{\ot}{\otimes}

\newcommand{\otte}{\otimes_\theta}

\date{March 5, 2015}

\begin{document}
\maketitle

\begin{abstract}
	Fully Poincar{\'e} covariant quantum field theories on non-commutative Moyal Minkowski spacetime so far have been considered in their vacuum representations, i.e. at zero temperature. Here we report on work in progress regarding their thermal representations, corresponding to physical states at non-zero temperature, which turn out to be markedly different from both, thermal representations of quantum field theory on commutative Minkowski spacetime, and such representations of non-covariant quantum field theory on Moyal Minkowski space with a fixed deformation matrix.
\end{abstract}

\section{Introduction}

Quantum field theory on ``non-commutative spaces'' \cite{DouglasNekrasov:2001} has been pursued in the past years for a number of reasons: On the one hand, it is generally expected that our picture of spacetime as a ``classical'' Lorentzian manifold, with matter described by quantum fields that propagate on this spacetime, breaks down at extremely small scales of the order of the Planck length $\la_P\sim10^{-35}$m. One expects that localization measurements resolving such scales should not be possible \cite{DoplicherFredenhagenRoberts:1994}, and no sharp distinction between ``matter'' and ``geometry'' degrees of freedom should exist in this regime. Since a ``quantized'' space with non-commuting coordinates, say $X_1,...,X_{d-1}$ (modeled as self-adjoint elements of some ${}^*$-algebra) automatically yields uncertainty relations, i.e. lower bounds on products of uncertainties $\Delta_\om X_\mu\cdot\Delta_\om X_\nu$, QFT on non-commutative spaces naturally leads to models in which sharp joint measurements of 
several space-time coordinates is impossible.

Another motivation to consider such theories comes from renormalization theory: Since localization at sharp points is impossible in non-commutative spaces, one expects the ultraviolet divergences of QFT to be softer in such a setting than in usual ``commutative'' Minkowski space (see, for example, \cite{GrosseWulkenhaar:2005,Rivasseau:2007}). 

Non-commutative QFT models also play a prominent role in the spectral formulation of the standard model (see  \cite{ChamseddineConnes:1997_2,ChamseddineConnesMarcolli:2007_2,DevastatoLizziMartinetti:2014,ChamseddineConnesvanSuijlekom:2013_4} for just some sample articles), or in approaches to quantum gravity via non-commutative structures \cite{Steinacker:2007}.

Furthermore, also certain low-energy limits of string theory result in field theories QFT on non-commutative spaces \cite{SeibergWitten:1999,ConnesDouglasSchwarz:1998}, providing further motivation to study such models. 

Finally, one can also take the point of view that the study of QFT on non-commutative spaces can lead to new insights about QFT on commutative spaces \cite{BuchholzLechnerSummers:2011}.
\\\\
Independent of the motivation, there is one type of non-commutative structure which is by far the simplest and best studied one, namely Moyal-Weyl space. Here one postulates selfadjoint coordinate operators\footnote{For the sake of simplicity, we will stick to the physically most interesting case of $d=4$ spacetime dimensions. With minor modifications, our considerations also apply in general dimension $d\geq2$.} $X_0,X_1,X_2,X_3$, subject to the canonical commutation relations
\begin{equation}
	\left[ X_\mu, X_\nu \right] = i \,\theta_{\mu \nu}\cdot 1\,,
	\label{eqn_MWRels}
\end{equation}
where $1$ denotes the identity element of the algebra generated by the $X_\mu$, and the constants $\te_{\mu\nu}$ form a real skew symmetric $(4\times 4)$-matrix. Such commutation relations are, of course, well known from the Schrödinger representation of quantum mechanics. In that situation, the $X_\mu$ correspond to position and momentum operators, and $\te$ is given by a factor of $\hbar$ and the classical Poisson brackets. 

In the setting of Moyal-Weyl {\em spacetime}, however, the interpretation of the $X_\mu$ is that of spatial ($X_1,X_2,X_3$) respectively time ($X_0$) coordinates (instead of coordinates of phase space), defining a non-commutative version of Minkowski spacetime. The matrix $\te$ is a measure for the strength of non-commutative effects, and can be taken to be proportional to $\la_P^2$.

This new view of the commutation relations \eqref{eqn_MWRels} immediately leads to a well-known problem: The symmetries of \eqref{eqn_MWRels} do not match the Poincar\'e symmetry of ``classical'' Minkowski space. Whereas translations, implemented by $X_\mu\mapsto X_\mu+x_\mu\cdot 1$ with $x\in\Rl^4$, do leave \eqref{eqn_MWRels} invariant, the same is not true for Lorentz transformations $X_\mu\mapsto{\La_\mu}^\nu X_\nu$. Only those Lorentz transformations that fix the matrix $\te$ also respect the relation \eqref{eqn_MWRels}, thus breaking Lorentz symmetry down to a subgroup unless $\te=0$.

There are three different ways to deal with this situation, all of which have been considered in the literature. We recall them here and list a few sample publications for each approach.

First, one can simply accept the breaking of Lorentz symmetry, and work with models with less symmetry than their counterparts on commutative Minkowski space, see for example \cite{GrosseWulkenhaar:2004,LakhouaVignes-TourneretWallet:2007_2}. As a second option, one can modify the commutation relations \eqref{eqn_MWRels} in such a way that they become fully covariant. This point of view has in particular been put forward by Doplicher, Fredenhagen, and Roberts \cite{DoplicherFredenhagenRoberts:1995}, who replaced the constant matrix $\te$ by an operator-valued matrix $Q$ in the center of the algebra generated by the $X_\mu$, which has a full Lorentz orbit of real, Lorentz skew symmetric\footnote{Starting from this point on, we will use the Lorentz skew symmetric matrices         $\vartheta^\mu_\nu = \eta^{\mu \kappa} \theta_{\kappa \nu}$ instead of the Euclidean skew symmetric ones and again denote them by $\theta$.} $(4 \times 4)$-matrices,
\begin{align}\label{eq:Theta}
	\Theta=\{\La\te_0\La^{-1}\,:\,\La\in\LG\}\,,
\end{align}
as the joint spectrum of its components. Here $\theta_0$ serves to fix the particular orbit. See also \cite{GrosseLechner:2007,GrosseLechner:2008} for other covariant models. Third, using methods from quantum groups, one can ``twist'' the action of the Lorentz group in such a way that it is compatible with \eqref{eqn_MWRels} \cite{Zahn:2006,FioreWess:2007,ChaichianPresnajderTureanu:2005_2}. However, this approach is often equivalent to the previous one \cite{Piacitelli:2009-1}. 
\\
\\
In this article, we will consider the fully Poincar{\'e} covariant model proposed in \cite{GrosseLechner:2007}, with the aim to understand its behavior {\em at finite temperature}. As in the approach taken in \cite{DoplicherFredenhagenRoberts:1995}, this model involves a Lorentz orbit $\Theta$ \eqref{eq:Theta}, but differs in its localization properties. To briefly introduce the model (more details will be provided in the following section), consider the free scalar Klein-Gordon field $\phi$ on Fock space $\Hil$ (i.e., in its vacuum representation, at zero temperature), and coordinate operators $X_\mu$ satisfying \eqref{eqn_MWRels} for fixed~$\te$, realized on a ``coordinate Hilbert space'' $\VV$. On the tensor product space $\Hil\ot\VV$, consider the field
\begin{align}
	\phi^\ot_\te(x)
	=
	(2\pi)^{-2}\int dp\,\phiti(p)\ot e^{ip\cdot(X+x)}
	\,.
\end{align}
There exists a natural vacuum state on the algebra generated by the (smeared) field operators $\phi^\ot_\te$, and when one passes to its GNS representation, one finds an equivalent field operator $\phi_\te$, acting on Fock space $\Hil$ instead of $\Hil\ot\VV$ (see Section~\ref{sec:VacSector}). Concretely, this amounts to replacing the usual scalar free field $\phiti(p)$ (in momentum space) by its ``twisted'' version
\begin{align}\label{eq:deformed-field}
  \phiti_\te(p)
  =
  \phiti(p)\cdot e^{-\frac{i}{2} p \cdot \theta P}
  \,,
\end{align}
where $P=(P_0,P_1,P_2,P_3)$ denote the energy momentum operators on Fock space (see also \cite{BalachandranPinzulQureshiVaidya:2007}).

When understood in a proper distributional sense, the construction \eqref{eq:deformed-field} is unitarily equivalent to the one on $\cV \otimes \Hil$ for fixed $\theta$ \cite{GrosseLechner:2007}, but there is one big advantage: Fields for different $\theta$ are represented on the same Hilbert space and, since $\Hil$ is the usual Fock space of the Klein-Gordon field,  one has a unitary representation $U$ of the full Poincar{\'e} group. 
Adjoint action with $U(a,\Lambda)$ on a field operator $\phi_\theta$ yields $\phi_{\Lambda \theta \Lambda^{-1}}$ (which is compatible with \eqref{eqn_MWRels}) but more importantly, is another well-defined field operator
on the same Hilbert space $\Hil$. It is thus possible to make the theory covariant, by not just taking field operators $\phi_\theta$ for some fixed $\theta$, but instead $\phi_\theta$ for all $\theta$ from an orbit $\Theta$ \eqref{eq:Theta}. 

The model we want to consider is then described by the field algebra $\Pol_\Theta$ of all polynomials in (smeared) field operators $\phi_\te(f)$, where $\te$ runs over the Lorentz orbit $\Theta$, and $f$ over the space of all test functions $\Ss(\Rl^4)$. We also include the identity operator $1$, so that $\Pol_\Theta$ has the structure of a unital ${}^*$-algebra.
\\\\
In this contribution, we want to discuss thermal equilibrium states for this model as a concrete representative of a quantum field theory on a non-commutative spacetime. Below we list our motivations for this investigation and some of the questions we want to address.
\begin{description}
	\item[\bf Q1)] An important selection criterion for physically meaningful theories is their ability to allow for thermal states. ``Does the considered model have thermal equilibrium states?'', the basic question preceding any analysis of thermal behavior, will be investigated here. At first sight, the appearance of functions of the energy-momentum operators $P_\mu$ in the field algebra (already anticipated in the twisted field operators \eqref{eq:deformed-field}) seems to threaten this. Further investigation however shows that things are more complicated, warranting a more detailed study.
	\item[\bf Q2)] On scales accessible to us, QFT on non-commutative spaces can be expected to yield only tiny deviations from the predictions of usual QFT. However, it is not very clear what these deviations are, or how they could be observed. In the model at hand, the effect of the noncommutativity parameter $\te$ on scattering processes (at zero temperature) was investigated in \cite{GrosseLechner:2007}, and shown to lead to modifications in the phase shift of two-particle scattering. Here we ask in the same model ``what are the observational consequences of a non-commutative structure of spacetime on the thermal behavior?''  
	\item[\bf Q3)] The model under consideration can also be considered as a ``deformed'' theory of fields on ordinary Minkowski spacetime, where the fields are however only localized in spacelike wedge regions. This observation is in line with several investigations of deformations of quantum field theories which are compatible with Poincar\'e covariance and (at least partial) localization/causality properties \cite[Sect.~6]{Summers:2011},\cite{Lechner:2012}. It is a common theme of these approaches that they make use of a condition of positivity of the energy. Such a spectrum condition holds in vacuum representations of QFT, where the vacuum state is a ground state for the energy, but not in thermal representations because of the presence of a heat bath from which arbitrary amounts of energy can be extracted. In fact, typical deformation schemes do not lead to wedge-locality when considering thermal field theory instead of QFT in its vacuum representation \cite{Morfa-Morales:2012}. 
  
	From the point of view of deformation theory, it is interesting to ask what one can do without the spectrum condition. As we shall see, an analysis of thermal representations of deformed vacuum field theories, such as the one considered here, can shed new light onto this question.
\end{description}

On a mathematical level, these questions turn out to be best analyzed with {\em algebraic methods}. As appropriate for proceedings of a conference covering a wide range of topics, we will next outline the tools we rely on, essentially reviewing the formalism of KMS states on unital ${}^*$-algebras (see, for example, \cite{BratteliRobinson:1997}).

Following standard procedure in quantum physics \cite{Emch:1972}, {\em states} on the field (or observable) algebra are modeled as certain functionals $\om:\Pol_\Theta\to\Cl$, mapping field polynomials, generically denoted $F\in\Pol_\Theta$, to their expectation values $\om(F)$. These functionals are taken to be linear, normalized in the sense that $\om(1)=1$, and {\em positive}. The precise definition of positivity depends a little on the topological structure of the algebra\footnote{Usually, one also requires states to be continuous. In the case of a $C^*$-algebra, continuity is a consequence of positivity, but for more general topological algebras such as the field algebra considered here, this is not the case. For the sake of simplicity, we do not go into these details here.}, but always involves the condition that ``squares'' $F^*F$ should have positive expectation values,
\begin{align}\label{eq:Positivity}
	\om(F^*F)\geq0\,,\qquad F\in\Pol_\Theta\,,
\end{align}
which is necessary for $\om$ to have a probability interpretation (positive uncertainties). 

In our model, a simple example of such a state is the vacuum state on $\Pol_\Theta$: As $\Pol_\Theta$ consists of operators on the Fock space $\Hil$ containing the (normalized) vacuum vector $\Om$, the vacuum state is $\om_{\rm vac}(F)=\langle\Om,F\Om\rangle$. 

However, the concept of state is not restricted to vector states, and thinking of thermal equilibrium states in quantum mechanics, one might consider states of the Gibbs form $\om_\beta(F)=Z^{-1}{\rm Tr}(F\,e^{-\beta H})$, where $H=P_0$ is the Hamiltonian, $\beta=1/T$ the inverse temperature, and $Z={\rm Tr}(e^{-\beta H})$ the partition function. Such Gibbs ensembles are well-defined in finite volume, where the Hamiltonian has discrete spectrum and the trace $Z$ exists. In the case of a theory in infinite volume, such as the one considered here, Gibbs states do however not exist per se, but rather only as quantities approximating the equilibrium states in infinite volume (thermodynamic limit). 

It is long since known \cite{HaagHugenholtzWinnink:1967} that a more intrinsic formulation of thermodynamic equilibrium states can be given in terms of the Kubo-Martin-Schwinger (KMS) condition. This condition trades the explicit Gibbs form of the state for a boundary condition, involving analytic continuation along the Heisenberg dynamics
\begin{align}
	\tau_t(F):=e^{itP_0}Fe^{-itP_0}
	\,.
\end{align}
The precise definition of the KMS property again depends a little on the topological structure of the algebra \cite{FredenhagenLindner:2013}, but always involves the following condition (which is sufficient in the case of a $C^*$-algebra, and to which we restrict here for the sake of simplicity).
A {\em KMS state at inverse temperature $\beta>0$} is a state $\om$ on $\Pol_\Theta$ which has the property that for each two field polynomials $F,G\in\Pol_\Theta$, the function 
\begin{align}
	f_{F,G}(t):=\om(F\tau_t(G))\,,\qquad t\in\Rl\,,
\end{align}
has an analytic continuation to the complex strip $\Rl+i(0,\beta)$, is bounded and continuous on the closed strip $\Rl+i[0,\beta]$, and satisfies the boundary condition
\begin{align}\label{eq:KMS-Condition}
	f_{F,G}(t+i\beta)
	=
	\omega(\tau_t(G)F)
	\,,\qquad t\in\Rl.
\end{align}
KMS states show the typical features of thermal equilibrium states in general, such as invariance under the dynamics, stability, and passivity \cite{BratteliRobinson:1997}. For states on ``type I von Neumann factors'' \cite{BratteliRobinson:1997}, as they are encountered in quantum mechanics, and dynamics with a Hamiltonian $H$ such that ${\rm Tr}(e^{-\beta H})$ exists, the KMS states coincide with the Gibbs states. Thus the KMS condition provides a good characterization of thermal equilibrium.
\\
\\
Returning to our above list of questions, the first question Q1) can therefore be phrased as the question about existence of KMS states on $\Pol_\Theta$. Explicit knowledge of such states (say, in terms of its $n$-point functions) is also the basis of a quantitative understanding of the thermodynamic behavior of the considered system (question Q2)). In the present conference paper, we restrict ourselves to announce several results concerning the algebraic structure of $\Pol_\Theta$ and its equilibrium states, including explicit formulas for their $n$-point functions. A full analysis with detailed proofs will be presented elsewhere. We however wish to motivate why it is useful to have a closer look at the algebraic structure of $\Pol_\Theta$ in this context, and to this end, proceed with an outline of the paper.
\\\\
In Section~\ref{sec:VacSector}, we introduce the vacuum representation of the model in more detail. The algebra $\Pol_\Theta$, containing smeared polynomials in all field operators $\phi_\te$, $\te\in\Theta$, obviously has as subalgebras the polynomials in fields $\phi_\te$ with a {\em fixed} $\te\in\Theta$. These subalgebras will be denoted $\Pol_{\{\te\}}\subset\Pol_\Theta$, and referred to as ``fibers''. Since each such fiber is invariant under the dynamics $\tau_t$, it is clear that given a KMS state on $\Pol_\Theta$, it restricts to KMS states on each of the fibers $\Pol_{\{\te\}}$. Thus, as a preliminary step, we consider in Section~\ref{sec:SingleFibre} the KMS states on a fixed fiber. 

It turns out that the thermal states of $\Pol_{\{\te\}}$ can be easily analyzed completely. Namely, by exploiting a natural bijection with the ``zero fiber'' $\Pol_{\{0\}}$, containing the polynomials in the undeformed free Klein-Gordon field, we find that the KMS states on $\Pol_{\{\te\}}$ and $\Pol_{\{0\}}$ are in one to one correspondence. Moreover, one can explicitly calculate all $n$-point functions of these equilibrium states by basically the same techniques as in the undeformed case. These results are obtained by making use of the machinery underlying Rieffel's deformation theory \cite{Rieffel:1992} and the related concept of warped convolutions \cite{BuchholzSummers:2008,BuchholzLechnerSummers:2011}, and allow us to substantially generalize findings of Balachandran and Govindarajan about the thermal four-point function at fixed noncommutativity $\te$ \cite{BalachandranGovindarajan:2010}.

Having settled the structure of thermal equilibrium states at fixed $\te$ (i.e., for the non-covariant models build on \eqref{eqn_MWRels}), we proceed to the analysis of thermal equilibrium states on the full field algebra $\Pol_\Theta$ in Section~\ref{sec:FullAlg}. Here a completely new aspect enters, both on the level of the structure of the field algebra, and on the level of its thermal equilibrium states. To see this effect, note that the field operators\footnote{Here and in the following, we adopt the convention to rescale $\te$ by a factor $\frac{1}{2}$ to obtain simpler formulas.} \eqref{eq:deformed-field} satisfy the twisted commutation relation
\begin{align}
	\phiti_\te(p)\phiti_{\te'}(p')-e^{ip\cdot(\te+\te')p'}\,\phiti_{\te'}(p')\phiti_{\te}(p)
	=
	\tilde{C}(p,p')\,e^{ip\cdot(\te-\te')P}
	\,,
\end{align}
where $\tilde{C}(p,p')$ denotes the (number-valued) momentum space commutator of two undeformed fields $\phiti(p),\phiti(p')$. For $\te,\te'\in\Theta$, $\te\neq\te'$, this shows that certain functions of the energy momentum operators $P_\mu$ are elements of $\Pol_\Theta$ (which are not present in the fibers $\Pol_{\{\te\}}$). In addition to the fibers $\cP_{\{\theta\}}$, the algebra thus also contains a commutative subalgebra $\cT_\Theta$ containing functions of $P$. 

This subalgebra has the peculiar property that, consisting of functions of the (commuting) energy momentum operators $P_0,...,P_3$, each of its elements is invariant under the dynamics $\tau_t(F)=e^{itP_0}Fe^{-itP_0}$. 

Thus the functions $f_{F,G}(t)=\om(F\tau_t(G))$ appearing in the KMS condition are {\em constant} for each $F,G\in\T_\Theta$, and because $\T_\Theta$ is abelian, $\om$ automatically satisfies the required analyticity and boundary condition \eqref{eq:KMS-Condition}. This observation implies that, in sharp contrast to the situation on the fibers $\Pol_{\{\te\}}$, {\em each} state on $\T_\Theta$ is a KMS state. In other words, the KMS condition does not put any constraints on the expectation values $\om(F)$, $F\in\T_\Theta$.

As a consequence, a very large family of KMS functionals (at fixed temperature) appears. Usually, the appearance of several distinct KMS states at the same temperature is taken as a sign for different thermodynamic phases that can coexist at that temperature. The picture of many ``non-commutative phases'' appearing at any finite temperature seems to be hard to maintain, and in fact, it has to be taken into account that these KMS functionals must also meet the essential positivity condition \eqref{eq:Positivity} in order to have physical meaning. While on the level of linear relations used to determine $n$-point functions, there are many similarities to thermal states of free field models --- with the exception of the mentioned additional freedom on $\T_\Theta$ --- things get more complicated for the nonlinear relation \eqref{eq:Positivity}. 

Using yet another subalgebra $\J_\Theta\subset\Pol_\Theta$, namely an ideal generated by the projection onto the vacuum vector, the positivity questions can be completely settled. It turns out that the positivity requirement singles out a {\em unique} KMS state on $\Pol_\Theta$. We discuss some of its properties in Section~\ref{sec:FullAlg}. In particular, we describe its associated GNS representation, which is ``more commutative'' than the algebra $\Pol_\Theta$ at temperature zero. We end by explaining the implications of this observation for the question Q3) on our list, and with an account of some open questions which we are currently investigating.

\section{The model}\label{sec:VacSector}

In this section we define the vacuum representation of the model that we consider.

Fixing a mass parameter $m>0$, we consider the single-particle Hilbert space $$\Hil_1 := L^2(\Rl^4, d\mu_m(p)),$$ with the mass shell measure
$d \mu_m(p) = \delta(p^0 - \epsilon(\V{p}))\, \frac{d^4p}{2 \epsilon(p)}$, where
$p = (p^0, \V{p})$ and $\epsilon(\V{p}) = (m^2 + \V{p}^2)^{1/2}$ is the single particle Hamiltonian. On $\Hil_1$ there acts the unitary irreducible representation $U_1$ of the proper orthochronous
Poincar{\'e} group with mass $m$ and spin zero, given by $(U_1(a,\La)\psi)(p)=e^{ip\cdot a}\psi(\La^{-1}p)$.

The full vacuum Hilbert space $\Hil$ is defined as the Bose Fock space over
$\Hil_1$. It carries the second quantized representation $U$ of the 
Poincar{\'e} group and contains the $U$-invariant vacuum vector $\Omega$.
For pure translation operators, we will also write
\begin{align}\label{eq:Translations}
	U(x)=e^{ix\cdot P}\equiv U(x,1)\,,\qquad x\in\Rl^4\,.
\end{align}
Here $P=(P_0,P_1,P_2,P_3)$ are the energy-momentum operators (with joint spectrum contained in the forward light cone). In particular, $P_0=H$ is the Hamiltonian giving the dynamics with respect to which we will be looking for thermal equilibrium.

The adjoint action of $U$ on operators will be denoted $\alpha_{x,\Lambda}(F)=U(x,\La)FU(x,\La)^{-1}$, and for the dynamics (time translations), we reserve the shorter notation
\begin{align}\label{eq:tau}
	\tau_t(F):=e^{itH}Fe^{-itH}=\alpha_{(t,0,0,0),1}(F)\,.
\end{align}
On $\Hil$, the (undeformed) free scalar Klein-Gordon field $\phi$ of mass $m$ acts in the 
usual manner, i.e.
\[
  \phi(f) = a^\dagger(\tilde{f}) + a(\overline{\tilde{f}})
\]
with $a$, $a^\dagger$ Bose creation and annihilation operators, 
$f \in \Schw(\Rl^4)$ a test-function from Schwartz space and 
$f \mapsto \tilde{f}$ the Fourier transform. We will often write $\phi$ in terms of its distributional kernels in momentum space, i.e. 
\begin{align}
	\phi(f)=\int d^4p\,\phiti(p)\fti(-p)\,.
\end{align}
Well-known properties of $\phi$ that are relevant here are first of all its c-number commutation relations
\begin{align}\label{eq:UndeformedFieldCommutator}
	[\phiti(p),\phiti(q)]
	=
	\delta(p+q)\,\epsilon(p^0)\,\delta(p^2+m^2)\cdot 1
	=:\Cti(p,q)\cdot 1
	\,,
\end{align}
which imply in particular {\em locality}, that is $[\phi(x),\phi(y)]=0$ for $x$ spacelike to $y$. 

Furthermore, $\phi$ transforms covariantly under the representation $U$, i.e. 
\begin{align}\label{eq:phi-covariance}
	U(a,\La)\phiti(p)U(a,\La)^{-1}
	=
	e^{ip\cdot a}\cdot\phiti(\La p)\,.
\end{align}

According to the terminology used in the Introduction, the field algebra of the free field $\phi$ is denoted $\Pol_{\{0\}}$, because it corresponds to fixed deformation parameter $\te=0$. Since we are interested in QFT at finite temperature, we also recall the known results about the KMS states of $\Pol_{\{0\}}$.

\begin{Lem}\label{Lem:KMSonUndeformedFiber}
	\begin{enumerate}[a)]
		\item There exists a unique\footnote{``Unique'' refers to fixed inverse temperature $\beta>0$, and fixed dynamics $\tau$ \eqref{eq:tau}. Furthermore, for the ${}^*$-algebraic setting considered here, some weak continuity 
                          assumptions are required in order to have analytical 
                          tools like Fourier transforms available. Here and in 
                          the following we will always restrict to states whose
                          $n$-point functions are temperate distributions.}
                      KMS state $\omega_0$ on $\cP_{\{0\}}$.
		\item $\omega_0$ is translationally invariant, i.e. 
                      $\omega_0\circ\alpha_{x,1}=\omega_0$ for all $x\in\Rl^4$.
		\item \label{enu:KMSfun} If $\varphi$ is a KMS functional\footnote{That is, a linear functional which satisfies the KMS condition, but is not necessarily normalized or positive.} on 
                      $\cP_{\{0\}}$, then there exists $c\in\Cl$ such that 
                      $\varphi=c\cdot\omega_0$.
	\end{enumerate}
\end{Lem}
Item \ref{enu:KMSfun}) tells us that the linear relations
already fix the state up to normalization, so the problem of determining
KMS states is essentially a linear one. 

Explicitly, $\omega_0$ can be described by its $n$-point functions, which are 
of quasi-free form. Denoting the distributional 
kernels of $f_1\otimes \ldots \otimes f_n\mapsto\omega_0(\phi(f_1)\cdots\phi(f_n))$ (in momentum space) by $\tilde{\omega}_{0,n}(p_1,...,p_n)$, we have the 
formulas, $n\in\N_0$,
\begin{align}\label{eq:om0-n-point-functions-odd}
	\tilde{\omega}_{0,2n+1}(p_1,...,p_{2n+1})
	&=
	0
	,\\
	\tilde{\omega}_{0,2n}(p_1,...,p_{2n})
	&=
	\sum_{(\V{l},\V{r})}
	\prod_{j=1}^n
	\tilde{\omega}_{0,2}(p_{l_j},p_{r_j})
	& 
	\tilde{\omega}_{0,2}(p,q)
	&=
	\frac{\tilde{C}(p,q)}{1-e^{\beta p^0}}.
	\label{eq:om0-n-point-functions-even}
\end{align}
Here the sum runs over all {\em contractions} $(\V{l},\V{r})$ of 
$\{1,...,2n\}$, i.e. sets of ordered multi indices 
$\V{l}=(l_1,...,l_n)$, $\V{r}=(r_1,...,r_n)$ such that 
$l_1<...<l_n$, $l_k<r_k$, $k=1,...,n$, and 
$\{l_1,...,l_n,r_1,...,r_n\}=\{1,...,2n\}$.

Regarding b), recall that any KMS state is invariant under the time translations $\tau$ it refers to, but not necessarily invariant under spatial translations as well. It is clear from \eqref{eq:om0-n-point-functions-even} and \eqref{eq:UndeformedFieldCommutator} that $\om_0$ is also invariant under spatial rotations (but not under Lorentz boosts). However, the only relevant symmetry of $\om_0$ that we will exploit later is its invariance under spacetime translations.

We also recall the GNS representation of $\Pol_{\{0\}}$ with respect to its KMS state $\om_0$, the well-known \label{GNS}
Araki-Woods representation \cite{ArakiWoods:1966}: The representation Hilbert space is\footnote{$\overline{\Hil}$ denotes the conjugate of the vacuum Fock space $\Hil$.}   $\Hil_0:=\Hil\ot\overline{\Hil}$, the vector implementing the KMS state is $\Om_0:=\Om\ot\overline{\Om}$, and the vacuum fields $\phi(f)\in\Pol_{\{0\}}$ are represented on $\Hil_0$ as $\pi_0(\phi(f))=\phi((1+\rho)^{1/2}f)\ot\overline{1}+1\ot\overline{\phi(\rho^{1/2}f)}$, where $\rho$ acts in momentum space by multiplying with $p\mapsto(e^{\beta\epsilon(\V{p})}-1)^{-1}$. These data form the GNS triple of $(\Pol_{\{0\}},\om_0)$, i.e.
\begin{align}
	\om_0(F_0)=\langle\Om_0,\pi_0(F_0)\Om_0\rangle_{\Hil_0}\,,\qquad F_0\in \Pol_{\{0\}}\,.
\end{align}
As $\om_0$ is invariant under translations, $\Hil_0$ also carries a unitary representation $U_0$ of the space-time translations, related to the translations $U$ in the vacuum representation by $U_0(x)=U(x)\ot\overline{U(x)}$. Because of the conjugation in the second tensor factor, this representation does not have positive energy, but rather all of $\Rl^4$ as its energy-momentum spectrum.
\\
\\
Having recalled the relevant structures of the undeformed free field theory, we now proceed to its deformed version. Let $\te$ denote a real Lorentz skew symmetric $(4\times 4)$-matrix, and define
\begin{align}\label{eq:PhiTheta-distributional}
	\phiti_\te(p):=\phiti(p)\,U(-\te p)\,.
\end{align}
Since $U(0)=1$, this definition returns the original field $\phi_0=\phi$ for $\te=0$.

The definition \eqref{eq:PhiTheta-distributional} has to be understood in the sense of distributions. For a proper definition in terms of test functions, consider the expressions
\[
  (f_1, \ldots, f_n) 
    \mapsto 
	\Psi_n(f_1 \otimes \ldots \otimes f_n)
    := \phi(f_1) \cdots \phi(f_n) \Omega
    \,,
\]
which define vector-valued distributions $\Psi_n: \Schw(\Rl^{4n}) \to \Hil$, $n \in \N_0$, and the linear span $\cD$ of all their ranges is a dense
subspace of $\Hil$ \cite{StreaterWightman:1964}. In particular, $\Psi_n$ extends to all test functions of $4n$ variables, also those which are not of product form. The deformed field operator $\phi_\te(f)$, smeared with a test function $f\in\Ss(\Rl^4)$, can then be defined on~$\cD$ as 
\cite{Soloviev:2008,GrosseLechner:2008}
\begin{equation}
\label{eqn:DefFieldOp}
\phi_\theta(f) \Psi_n(g_n) 
  := 
\Psi_{n+1}(f \otimes_\theta g_n), 
  \qquad
 g_n\in \Schw(\Rl^{4n}),\, n \in \N_0,
\end{equation}
where the twisted tensor product $f \otimes_\te g_n$ is given in Fourier space by
\begin{align}\label{eq:WeylMoyalTensorProductI}
   (\widetilde{f \otimes_\theta g_n})(p, q_1, \ldots, q_n)
     :=
   \prod_{r=1}^n
     e^{i p \cdot \theta q_r} \tilde{f}(p) \cdot
                                \gti_n(q_1, \ldots, q_n).
\end{align}

The properties of $\phi$ are affected by the deformation $\phi\leadsto\phi_\te$. Regarding covariance, one finds by straightforward calculation the modified transformation law
\begin{align}\label{eq:PhiTheta-Symmetry}
	U(a,\La)\phiti_\te(p)U(a,\La)^{-1}
	=
	e^{ia\cdot p}\,\phiti_{\La\te\La^{-1}}(\La p)\,,
\end{align}
which states that translations are still a symmetry of the deformed fields at fixed $\te$, but Lorentz transformations in general change the deformation parameter (cf. the discussion of the Weyl-Moyal relations in the Introduction). Also note that $\phi_\te$ is not local if $\te\neq0$. However, certain remnants of locality are still present after the deformation. Comparing $\phi_\te$ with the inversely deformed field $\phi_{-\te}$, one finds localization in Rindler wedges. We refer to \cite{GrosseLechner:2007} for more details on this, and the consequences for two-particle scattering.

Denoting the algebra of all polynomials in (smeared) field operators $\phi_\te(f)$ (with fixed~$\te$) by $\Pol_{\{\te\}}$, the transformation law \eqref{eq:PhiTheta-Symmetry} yields
\begin{align}
	\alpha_{x,\La}(\Pol_{\{\te\}})
	=
	\Pol_{\{\La\te\La^{-1}\}}\,.
\end{align}
We therefore enlarge our field algebra to contain $\Pol_{\{\La\te\La^{-1}\}}$ for all Lorentz transformations~$\La$ in order to arrive at a fully covariant model theory.

\begin{Defn}
	Let $\Theta = \{ \Lambda \theta_0 \Lambda^{-1} \mid \Lambda \in \LGpo \}$ be the
	orbit of some reference matrix $\theta_0\in\Rl_-^{4\times 4}$ under the Lorentz group~$\LGpo$. Then $\cP_\Theta$ is defined
	as the ${}^*$-algebra generated by all $\Pol_{\{\te\}}$, $\te\in\Theta$.
	The ${}^*$-subalgebras $\cP_{\{\theta\}} \subset \Pol_\Theta$, $\theta \in \Theta$,
	will be referred to as the fibers (over~$\theta$) of $\cP_\Theta$.
\end{Defn}

This ${}^*$-algebra of unbounded operators is the starting point for our analysis of thermal states in the following sections.

\section{KMS states for the fiber algebras}\label{sec:SingleFibre}

Given any KMS state on the full field algebra $\Pol_\Theta$, one obtains KMS states on all of its fibers $\Pol_{\{\te\}}\subset\Pol_\Theta$, $\te\in\Theta$, by restriction. In this section, we therefore consider the KMS states of the fiber algebras $\Pol_{\{\te\}}$ as a first step towards understanding the KMS states of~$\Pol_\Theta$. This amounts to determining the thermal equilibrium states of QFT models without Lorentz covariance, formulated on Moyal space with fixed $\te$.

The analysis of thermal states on $\cP_{\{\theta\}}$ becomes most transparent
if one uses the fact that $\cP_{\{\theta\}}$ can be seen as a {\em deformation} of the ``zero fiber'' $\cP_{\{0\}}$ (the field algebra of the Klein Gordon field on commutative Minkowski spacetime), related to introducing a new product on $\Pol_{\{0\}}$.

This situation is familiar from deformation quantization, where one introduces (non-commutative) star products on the (initially commutative) algebra of classical observables by pulling back the operator product of the quantum observables. In fact, since the commutation relations \eqref{eqn_MWRels} are just the Heisenberg commutation relations, it is no surprise that there are several mathematical analogies between theories on Moyal space with a fixed non-commutativity matrix $\te$ on the one hand, and quantization on the other hand. In the present setting, where also the algebra $\Pol_{\{0\}}$ corresponding to the undeformed situation is non-commutative, the right tools for setting up such a deformation are Rieffel's procedure \cite{Rieffel:1992} and its generalizations, in particular the ``warped convolution'' \cite{BuchholzSummers:2008,BuchholzLechnerSummers:2011}. We briefly recall here the relevant structures.

To begin with, we introduce the ``Weyl-Moyal tensor product'' between test functions of several variables as a slight generalization of \eqref{eq:WeylMoyalTensorProductI}. For $f_n\in\Ss(\Rl^{4n})$, $g_m\in\Ss(\Rl^{4m})$, we define 
\begin{align}\label{eq:WeylMoyalTensorProductII}
	\widetilde{(f_n\ot_\te g_m)}(p_1,..,p_n;q_1,...,q_m)
	&:=
	\prod_{l=1}^n\prod_{r=1}^m e^{ip_l\cdot\te q_r}\cdot\fti_n(p_1,...,p_n)\,\gti_m(q_1,...,q_m)\,.
\end{align}
This associative product on the tensor algebra over $\Ss(\Rl^4)$ induces a new, ``deformed'' product $\times_\te$ between field operators,
\begin{align}
	&\left(\int d^{4n}x\, f_n(x_1,...,x_n)\,\phi(x_1)\cdots\phi(x_n)\right)
	\times_\te
	\left(\int d^{4m}y\, g_n(y_1,...,y_m)\,\phi(y_1)\cdots\phi(y_m)\right)
	\nonumber
	\\
	&
	:=
	\int d^{4n}x\int d^{4m}y\, (f_n\otte g_m)(x_1,...,x_n;y_1,...,y_m)\,\phi(x_1)\cdots\phi(x_n)\phi(y_1)\cdots\phi(y_m)
	\,.
	\nonumber
\end{align}
This product is an adaptation of Rieffel's product \cite{Rieffel:1992} to field algebras as they appear in Wightman QFT \cite{GrosseLechner:2008}. Equipping the zero fiber $\Pol_{\{0\}}$ with the product $\times_\te$ instead of its original product (the usual operator product, which for field operators corresponds to the normal tensor product between test functions) results in a deformed version of $\Pol_{\{0\}}$: The linear structure is the same as that of $\Pol_{\{0\}}$, and thanks to $(F_0\times_\te G_0)^*=G_0^*\times_\te F_0^*$ and $F_0\times_\te1=1\times_\te F_0=F_0$, $F_0,G_0\in\Pol_{\{0\}}$, also the identity and star involution of $\Pol_{\{0\}}$ are unchanged \cite{Rieffel:1992,GrosseLechner:2008}. We write $(\Pol_{\{0\}},\times_\te)$ when we want to emphasize that we consider the vector space $\Pol_{\{0\}}$ equipped with the product $\times_\te$.

To formulate the relation between $(\Pol_{\{0\}},\times_\te)$ and the fiber $\Pol_{\{\te\}}$ over $\te\in\Rl^{4\times 4}_-$, we introduce the ``warped convolution'' map 
$D_\theta : \cP_{\{0\}}\to\cP_{\{\theta\}}$, which is defined on the generating 
fields by 
\begin{align}\label{eq:DeformationMap}
	D_\theta(\phi(f_n))
	&:=
	\phi_\theta(f_n)
	,\qquad
	f_n\in\Schw(\Rl^{4 n})
	,
\end{align}
and then extended by linearity. 

\begin{Prop}\label{Prop:DeformationMap}
	The map $D_\theta$ is an isomorphism of unital ${}^*$-algebras between the zero fiber $(\Pol_{\{0\}},\times_\te)$, with the product $\times_\te$, and the fiber $\Pol_{\{\te\}}$ over $\te$, with the usual operator product. That is, $D_\te$ is a linear bijection satisfying for     $F_0,G_0\in\cP_{\{0\}}$,
	\begin{enumerate}[a)]
		\item \label{enu:warpprod} $D_\theta(F_0 \times_\theta G_0)=D_\theta(F_0)D_\theta(G_0)$.
		\item $D_\theta(1)=1$.
		\item $D_\theta(F_0^*)=D_\theta(F_0)^*$.
		
	\end{enumerate}
	Moreover, $D_\te$ commutes with translations, i.e.
	\begin{align}
		\alpha_{x,1}\circ D_\te=
		D_\te\circ\alpha_{x,1}\,,\qquad x\in\Rl^4\,.
	\end{align}
\end{Prop}

We mention as an aside that just like $\Pol_{\{0\}}$, also $\Pol_{\{\te\}}=D_\te(\Pol_{\{0\}})$ has the vacuum vector $\Om$ as a cyclic vector because $D_\te(F_0)\Om=F_0\Om$, $F_0\in\Pol_{\{0\}}$ \cite{BuchholzLechnerSummers:2011}.

Since $(\Pol_{\{0\}},\times_\te) $ and $\Pol_{\{0\}}$ coincide as vector spaces, it is straightforward to pass from linear\footnote{All functionals considered here are linear, so that we refer to them just as functionals for brevity.} functionals on $\Pol_{\{0\}}$ to functionals on $(\Pol_{\{0\}},\times_\te)$: One can simply use the {\em same} functional. Formulated in terms of $\Pol_{\{\te\}}=D_\te(\Pol_{\{0\}})$, this means that a functional $\om:\Pol_{\{0\}}\to\Cl$ is transported to a functional $\om_\te:\Pol_{\{\te\}}\to\Cl$, defined as
\begin{align}\label{eq:DeformedState}
	\omega_\theta
	:=
	\omega_0\circ {D_\theta}^{-1}\,.
\end{align}
This definition yields a one to one correspondence between functionals on the undeformed fiber $\Pol_{\{0\}}$ and the deformed fiber $\Pol_{\{\te\}}$. For $\te=0$, one recovers the original functional $\om_0=\om$.

In view of Proposition~\ref{Prop:DeformationMap}, the correspondence \eqref{eq:DeformedState} preserves normalization and invariance under translations, i.e. $\om_\te(1)=1$ if and only if $\om_0(1)=1$, and $\om_\te\circ\alpha_{x,1}=\om_\te$ for all $x\in\Rl^4$ if and only if $\om_0\circ\alpha_{x,1}=\om_0$ for all $x\in\Rl^4$.

We are interested in KMS states, and therefore also have to consider the compatibility of the map $\om\mapsto\om_\te$ with positivity, and with the KMS condition. The notion of positivity is a different one in $\Pol_{\{0\}}$ and $(\Pol_{\{0\}},\times_\te)$: Whereas squares of the form $F_0^*F_0$ are positive in the sense of $\Pol_{\{0\}}$, squares of the form $F_0^*\times_\te F_0$ are positive in the sense of $(\Pol_{\{0\}},\times_\te)$. These two positive cones do not coincide, and therefore a positive functional $\om$ on $\Pol_{\{0\}}$ is in general not positive on $(\Pol_{\{0\}},\times_\te)$ \cite{KaschekNeumaierWaldmann:2009}.

However, things simplify drastically when considering {\em translationally invariant} functionals $\om$ on $\Pol_{\{0\}}$. In this case, one has \cite{Soloviev:2008,GrosseLechner:2008}
\begin{align}\label{eq:translationallyinv}
	\om(F_0\times_\te G_0)
	=
	\om(F_0G_0)\,,\qquad
	F_0,G_0\in\Pol_{\{0\}}\,,
\end{align}
similar to the well-known formula $\int d^4x\,(f\star_\te g)(x)=\int d^4x\,f(x)g(x)$ for the Weyl-Moyal star product. From \eqref{eq:translationallyinv} it is clear that with a translationally invariant, positive functional $\om:\Pol_{\{0\}}\to\Cl$, also $\om_\te:\Pol_{\{\te\}}\to\Cl$ is positive (at least on squares). Namely, for arbitrary $F_0\in\Pol_{\{0\}}$, we have
\begin{align*}
	\om_\te(D_\te(F_0)^*D_\te(F_0))
	=
	\om_\te(D_\te(F_0^*\times_\te F_0))
	=
	\om(F_0^*\times_\te F_0)
	=
	\om(F_0^*F_0)
	\geq0\,.
\end{align*}

Furthermore, by the same argument one also sees that the functions $$t\mapsto\om_\te(D_\te(F)\tau_t(D_\te(G)))=\om(F\tau_t(G))$$
are independent of $\te$, so that their analyticity properties and the relation \eqref{eq:KMS-Condition} between their boundary values, as required in the KMS condition, transfer from the undeformed fiber to the deformed one immediately.

All this applies only to KMS states that are invariant under all space-time translations. However, since the (unique) KMS state on $\Pol_{\{0\}}$ has this property (Lemma~\ref{Lem:KMSonUndeformedFiber}~b)), these considerations cover the situation of interest here. We find the following result.

\begin{Thm}\label{Theorem:KMSonFibers}
	\begin{enumerate}[a)]
		\item The mapping $\omega_0\mapsto\omega_\theta$ defined above 
                      is a  bijection between the functionals on $\cP_{\{0\}}$ 
                      and $\cP_{\{\theta\}}$. The functional $\omega_\theta$ is 
                      normalized (respectively translationally 
                      invariant) if and only if $\omega_0$ is normalized 
                      (respectively translationally invariant).
		\item For translationally invariant functionals $\omega_0$ on $\cP_{\{0\}}$, $\omega_\theta$ satisfies the KMS condition (is positive) if and only if $\omega_0$ satisfies the KMS condition (is positive).
		\item There exists a unique KMS state on each fiber 
                      $\cP_{\{\theta\}}$, given by 
                      $\omega_\theta=\omega_0\circ {D_\theta}^{-1}$.
        \item The $n$-point functions of this unique KMS state  $\omega_\theta$ are
		\begin{align}\label{eq:DeformedNPointFunctions-FixedTheta}
			\tilde{\omega}_{\theta,n}(p_1,...,p_n)
			:=
			\omega_\theta(\tilde{\phi}_\theta(p_1)\cdots
                                     \tilde{\phi}_\theta(p_n))
			=
			\prod_{1\leq l<r\leq n}e^{ip_l \cdot \theta p_r}
			\cdot\tilde{\omega}_{0,n}(p_1,...,p_n).
		\end{align}
	\end{enumerate}
\end{Thm}

The formula for $n$-point functions stated in part d) can be derived by straightforward calculation similar to the vacuum case \cite{GrosseLechner:2008}. 

The KMS states $\om_\te$ on $\Pol_{\{\te\}}$ can also be obtained by direct calculation of their $n$-point functions by imposing the KMS condition (cf. the calculation at the beginning of the next section). This approach was taken in \cite{Huber:2012}, see also \cite{BalachandranGovindarajan:2010} for a partial analysis in the context of non-covariant field theories on Moyal space. However, in our opinion, the more algebraic point of view taken here makes the situation particularly clear. Moreover, it applies without essential modifications also to more complicated field theories --- in particular, the commutation relations \eqref{eq:UndeformedFieldCommutator} are not needed in our approach, but often form the basis of direct calculations.

The situation described in Theorem~\ref{Theorem:KMSonFibers} closely resembles the case of undeformed fields:
For each inverse temperature $\beta>0$, there exists exactly one thermal equilibrium state, and this state is translationally
invariant. Furthermore, the distributional kernels of its $n$-point functions
in momentum space depend on $\te$ only via phase factors. Note however that due to these
phase factors, the thermal states on $\cP_{\{\theta\}}$ are no longer quasi-free.
\\
\\
This close relation between $\cP_{\{\theta\}}$ and $\cP_{\{0\}}$ is also  reflected on the level of the (GNS) representations $\pi_\theta$ and $\pi_0$ induced
by $\om_\te$ and $\om_0$: The GNS representation $\pi_\te$ of $\Pol_{\{\te\}}$ with respect to its unique KMS state $\om_\te=\om_0\circ{D_\te}^{-1}$ acts on the same representation space $\Hil_0$ as for $\te=0$, with the same implementing vector $\Om_0$ (cf. page~\pageref{GNS}). Analogously to the vacuum case \eqref{eq:deformed-field}, the representation of the deformed fiber $\Pol_{\{\te\}}$ can be realized as (in the sense of distributions)
\begin{align}
	\pi_\theta(\tilde{\phi}(p)) = \pi_\theta(\tilde{\phi}(p)) 
	U_0(-\theta p)
	\,.
\end{align}
That is, the thermal representation of a deformed fiber is given by warped convolution w.r.t. the translation representation $U_0$ on the thermal Hilbert space. In this sense, deforming and going to a thermal representation commute on a single fiber.

However, the algebra generated by the fields $\pi_\theta(\phi_\te(f))$, when $\theta$
is no longer fixed but ranges over the orbit $\Theta$, does \emph{not} form a representation of $\cP_\Theta$. This is so because of certain residual localization properties that are present in $\Pol_\Theta$: If the supports of two test functions $f,g$ are carefully chosen (in spacelike wedges depending on $\te$), then $[\phi_\te(f),\phi_{-\te}(g)]=0$ as a relation in $\Pol_\Theta$. This relation is a consequence of the spectrum condition in the vacuum case, and clearly persists in any representation of $\Pol_\Theta$. But the fields $\pi_\theta(\phi_\theta(f))$ and $\pi_{-\theta}(\phi_{-\theta}(g))$ do
not commute in general for $f$ and $g$ with the mentioned support properties. This is a consequence of the spectrum condition being violated in the thermal representation \cite{Morfa-Morales:2012}.

This observation already indicates that a thermal representation of the full algebra~$\cP_\Theta$ has to take place on a ``larger'' space than $\Hil_0$, such that more elements have the possibility to 
commute. How this representation, induced by a state
with many vanishing $n$-point functions, comes about, will be the topic of
the next and final section.

\section{\label{sec:FullAlg}KMS states for the covariant theory}

After discussing KMS states for the fiber subalgebras $\Pol_{\{\te\}}\subset\Pol_\Theta$ of our field algebra, we now come to the analysis of KMS states (and KMS functionals) on all of $\cP_\Theta$.
As announced in the Introduction, here we present some preliminary results
that will appear in more detail in a forthcoming publication.

The structure of thermal expectation values can most easily be described
by again using (idealized) Fourier transformed, deformed fields
$\tilde{\phi}_{\theta}(p)$ at sharp momenta $p$; some of the subtleties involved
in a mathematically completely rigorous discussion will be mentioned later.

To give a first impression of some of the new aspects that appear, we present the calculation of thermal two-point functions, the simplest non-vanishing $n$-point functions.  Using the symbolic expression 
$\tilde{\phi}_\theta(p) = \tilde{\phi}(p) U(-\theta p)$ \eqref{eq:deformed-field} for the deformed fields,
and the exchange relation $U(x)\phiti(p)=e^{ip\cdot x}\phiti(p)U(x)$ \eqref{eq:phi-covariance} between translation unitaries $U(x)$ and (undeformed) field operators $\tilde{\phi}(p)$, products $\phiti_{\te_1}(p_1)\cdots\phiti_{\te_n}(p_n)$ of deformed
fields can be written as products of undeformed fields, multiplied from
the right with a translation unitary and a phase factor. Assuming a KMS functional $\om$ on $\Pol_\Theta$, the task is thus to determine the expectation values of expressions like $\tilde{\phi}(p_1) \cdots \tilde{\phi}(p_n)
U(x)$ (where $x=-\sum_{j=1}^n\te_jp_j$ depends on the $\te_j$ and $p_j$) in $\om$. For the case of the two point function, this amounts to determining the expectation value of 
$\tilde{\phi}(p) \tilde{\phi}(p') U(-\theta p - \theta' p')$ from the KMS condition. 

On a formal level, this can be done as follows. Since the translation unitaries $U(x)$ are invariant under the dynamics, we have
\begin{align}
  \tau_t(\tilde{\phi}(p') U(x))
     & 
  = e^{i t p_0'} \tilde{\phi}(p') U(x) & 
  \Rightarrow\quad \label{eqn:KMSlhs}
  \tau_{t + i \beta}(\tilde{\phi}(p')) U(x)
     = &
  e^{i t p_0' - \beta p_0'}\,\tilde{\phi}(p') U(x) \,.
\end{align}
On the other hand, using the exchange relation of $U(x)$ with
the Fourier transformed fields, we find
\begin{equation}\label{eqn:KMSrhs}
   \tau_t (\tilde{\phi}(p') U(x)) \tilde{\phi}(p)
      =
   e^{i t p_0' + i x\cdot p} \tilde{\phi}(p') \tilde{\phi}(p) U(x)
      =
   e^{i t p_0' + i x\cdot p} \left(\tilde{C}(p', p) U(x)
                              + \tilde{\phi}(p) \tilde{\phi}(p') U(x) \right)
\end{equation}
Multiplying \eqref{eqn:KMSlhs} from the left with $\tilde{\phi}(p)$ and comparing with \eqref{eqn:KMSrhs}, we thus see that in a KMS functional $\om$ on $\Pol_\Theta$, the following equation must hold
\begin{align}
	 (e^{-\beta p_0' - i x\cdot p}-1) \omega(\tilde{\phi}(p) \tilde{\phi}(p') U(x))
     =
   \tilde{C}(p',p) \omega(U(x)) \,,
\end{align}
and, setting $x=-\te p-\te' p'$,
\begin{align}\label{eq:KMS2}
	\om(\phiti_\te(p)\phiti_{\te'}(p'))
	=
	\om(U(-\te p-\te' p'))\,\frac{\Cti(p,p')}{1-e^{\beta p^0+ip\cdot(\te p+\te'p')}}\,.
\end{align}
This calculation is, of course, only of a formal nature, a rigorous analysis requires in particular a proper treatment of the distributional nature of the appearing fields. Nonetheless, \eqref{eq:KMS2} illustrates the main difference to the KMS states on the fiber algebras, where $\te=\te'$ is fixed. In that case, the term $\omega(U(x))$ as well as the term $-i x\cdot p$ in the exponential function are not present, because of the translational invariance the commutator $\tilde{C}(p,p')$, leading to $p'=-p$. For $\te=\te'$, an analogous calculation directly determines the expressions for (the kernel of) the Fourier-transformed two-point function, up to normalization.

In the case of the full field algebra, however, we retain an expectation 
value of an element $U(x)$, which is precisely the point already mentioned in 
the Introduction: Functions of the (vacuum) energy-momentum operators appear in the algebra and 
their expectation values are a new freedom (not restricted by the KMS 
condition), which is reflected in the $n$-point functions. 

In other words, $\Pol_\Theta$ contains a commutative subalgebra $\cT_\Theta$ of functions of the momentum operators $P$, on which the dynamics acts trivially, and we also have to fix $\omega$ on this subalgebra. When working with properly smeared fields, $\T_\Theta$ can be identified with certain continuous functions on the joint spectrum of the momentum operators. It then follows that each  KMS-\emph{state} $\om$ on $\Pol_\Theta$ is given by an -- a priori arbitrary -- {\em measure} $\sigma$ 
(with mass bounded by one) on the energy-momentum spectrum.

By systematically exploiting the KMS condition along the same lines as in the
above calculation, the $n$-point functions can be determined for all $n\in\Nl$ \cite{Huber:2012}. These functions 
(which -- by forming linear combinations -- also give back the 
expectation values on $\cT_{\Theta}$), now contain as an additional 
ingredient the Fourier transform $\tilde{\sigma}$ of the measure $\sigma$. More precisely,
a modified version $\hat{\sigma}$ of $\tilde{\sigma}$, defined to take the value $1$ at the origin but otherwise identical to $\tilde{\sigma}$, appears and is formally related to the expectation values of translation operators
by $\omega(U(x)) = \hat{\sigma}(x)$. The $n$-point functions still have some structural similarity to quasi-free states: they vanish for an odd number of field operators, and for an even number are given by
	\begin{align}
		\omega(\tilde{\phi}_{\theta_1}(p_1)\cdots
                       \tilde{\phi}_{\theta_{2n}}(p_{2n}))
		&=
		\hat{\sigma}(-\sum_{j=1}^{2n}\theta_jp_j)
		\prod_{1\leq l<r\leq 2n}e^{ip_l\cdot\theta_l p_r}
		\sum_{(\V{l},\V{r})}\prod_{k=1}^n 
                                 \frac{\tilde{C}(p_{l_k},p_{r_k})}
                                      {1-e^{\beta p_{l_k}^0+ip_{l_k}\cdot
                                                \sum_{b=1}^{2n}\theta_b p_b}}
		\label{eqn:EvenNPointFunctions},
	\end{align}
where the sum runs over all contractions as in \eqref{eq:om0-n-point-functions-even}.

At this point a few remarks are in order.
\begin{itemize}
\item Whereas for the undeformed theory, and also for the Moyal-Weyl deformation
with fixed $\theta$, the thermal $n$-point functions for the Klein-Gordon field (with positive mass) are already uniquely determined
by the (linear) KMS-condition up to normalization, this condition is much less restrictive for the full field algebra~$\Pol_\Theta$: We have a huge additional freedom when only
using the KMS condition; this freedom is parametrized by Fourier transforms of distributions on the energy-momentum spectrum in the vacuum representation. Positivity on $\cT_\Theta$ reduces this freedom somewhat, but still all Fourier transforms $\tilde{\sigma}$
of measures on the spectrum remain as possible choices.
\item The structures appearing in these calculations look very similar to crossed products from the theory of operator algebras. Furthermore,
the calculation of KMS functionals on elements given as undeformed fields
times unitaries representing translations can be reinterpreted in terms of a ``twisted'' KMS condition \cite{BuchholzLongo:1999}. We hope to return to these interesting
links at a later point.
\item The formula \eqref{eqn:EvenNPointFunctions} should be read as follows: {\em If} a KMS functional $\om$ exists on $\Pol_\Theta$, then its $n$-point functions must have the specified form, with some measure $\sigma$. We are currently investigating also the converse direction, namely if given a measure $\sigma$ on the energy-momentum spectrum, \eqref{eqn:EvenNPointFunctions} {\em defines} a (KMS) functional on $\Pol_\Theta$. To answer this question, one must in particular make sure that the $n$-point functions \eqref{eqn:EvenNPointFunctions} yield a well-defined functional $\om$ on $\Pol_\Theta$. In the case of the undeformed theory,  the algebra $\Pol_{\{0\}}$ is simple and can be written as a quotient of a free algebra by a (maximal) ideal $\J_{\rm Rel}$. This ideal is generated by the usual relations a) linearity of $f\mapsto\phi(f)$, b) the Klein-Gordon equation, and c) the commutation relations \eqref{eq:UndeformedFieldCommutator}. To formulate a well-defined functional $\om$ on $\Pol_{\{0\}}$, one 
therefore only has to check that its $n$-point functions are consistent with a)--c). 

In contrast, $\Pol_\Theta$ is {\em not} simple. We will consider a non-trivial ideal in it below, but currently have no complete knowledge of its full ideal structure. Nonetheless, we expect that for any measure $\sigma$, \eqref{eqn:EvenNPointFunctions} defines a unique KMS functional on $\Pol_\Theta$.

We do however not make any claim towards positivity at this stage. Recall that in the case of a single fiber, each normalized KMS functional was automatically positive because of Lemma~\ref{Lem:KMSonUndeformedFiber}~c). This is not the case for the KMS functionals on the full field algebra, and we return to the positivity question below.
\item Although the structure of the $n$-point functions looks similar to the case 
of single fibers (and thus the non-deformed case), there are now terms 
involving all momenta instead of only pairs. Furthermore the simple 
Bose factors are modified by an additional imaginary term in the exponent. These two changes might appear as minor details at first sight, but  make it very hard to directly
check for positivity \eqref{eq:Positivity} of the $n$-point functions \eqref{eqn:EvenNPointFunctions}. Even simpler, necessary properties of states -- like hermiticity, i.e. $\om(F^*)=\overline{\om(F)}$ -- are not 
trivial to show.
\end{itemize}

The main question is now if there are choices for the measure 
$\sigma$ such that the above $n$-point functions determine a {\em positive} functional, i.e. a state on $\cP_\Theta$. To answer this question, it turns out to be most efficient to consider a particular ideal of $\cP_\Theta$, instead of directly analyzing the $n$-point functions. 

This ideal is defined as follows: A closer inspection of the subalgebra $\cT_{\Theta}\subset\Pol_\Theta$ shows that it also contains functions whose support intersects the energy-momentum spectrum only in $\{0\}$, i.e. the one-dimensional projection $E_\Om=|\Om\rangle\langle\Om|$ is an element of $\T_\Theta$. Multiplying this projection from both sides by elements from $\Pol_\Theta$, one generates an ideal
\begin{align}
	\J_\Theta
	:=
	\{FE_\Om G\,:\,F,G\in\Pol_\Theta\}
\end{align}
of finite-rank operators, which can be shown to be dense in the $C^*$-algebra of all compact operators on $\Hil$. 

By standard arguments \cite{BratteliRobinson:1997}, it follows that any KMS state $\om$ on $\Pol_\Theta$ must vanish identically on $\J_\Theta$. This implies directly that the measure $\sigma$ associated with $\om$ must satisfy $\sigma(\{0\})=0$. In fact, even more is true: By suitably choosing $F\in\Pol_\Theta$, one can show that $0=\om(FE_\Om)$ implies that the measure has to vanish altogether, $\sigma=0$. 

Thus all except one of the KMS functionals \eqref{eqn:EvenNPointFunctions}, namely the one given by the zero measure, are {\em not} positive. In contrast to the
undeformed or single fiber case, positivity plays an important 
role in selecting KMS-states: From an uncountable family of candidate functionals which satisfy the KMS-condition, positivity selects a \emph{unique} one.

Choosing $\sigma=0$ means that $\hat{\sigma}$ is only non-zero
at zero, which leads to a drastic decoupling of the individual
fibers: Only $n$-point functions in which the deformed fields $\phi_{\te_1},...,\phi_{\te_n}$ appear in
pairs belonging to the same fiber $\Pol_{\{\te_j\}}$ are non-zero. Furthermore, the subalgebra $\T_\Theta$ of functions of momentum operators is mapped to zero except
for the case of the identity operator; intuitively the unitaries ``average out'' by rapid oscillations in the thermodynamic limit.

To show that the remaining KMS functional, associated with the measure $\sigma=0$, {\em is} positive, it is best to directly specify its GNS representation, as we shall do next. The GNS space $\Hil_\Theta$ is a non-separable, infinite tensor 
product labeled by the elements in the orbit $\Theta$, and the GNS representation $\pi_\Theta$ reads (in the sense of distributions)
\begin{equation}
  \pi_\Theta(\phiti_\theta(p)) = \left(1 \otimes \cdots \otimes 1 \otimes 
                        \pi_0(\tilde{\phi}(p))\otimes 1 \otimes \cdots
                        \otimes 1 \right) \cdot
                        \bigotimes_{\vartheta \in \Theta} U_0(-\vartheta p)
\label{eqn:TheRepr}\,,
\end{equation}
where $\pi_0(\phiti(p))$ acts on the tensor factor labeled by $\te$, and $\pi_0$ is the GNS representation of the zero fiber w.r.t. its KMS state $\om_0$. Summing up, we have the following theorem.
\begin{Thm}
	On $\cP_\Theta$, for each $\beta>0$ there exists precisely one KMS state at inverse temperature $\beta$. Its $n$-point functions are
	\eqref{eqn:EvenNPointFunctions} with $\sigma = 0$, the thermal representation 
	it induces is given by \eqref{eqn:TheRepr}.
\end{Thm}

From the form of the GNS representation \eqref{eqn:TheRepr}, also the mentioned decoupling of the fibers is most transparent: For $\te\neq\te'$, one has the exchange relation
\begin{align}
	\pi_\Theta(\phiti_\te(p))\,\pi_\Theta(\phiti_{\te'}(p'))
	=
	e^{ip\cdot(\te+\te')p'}\,
	\pi_\Theta(\phiti_{\te'}(p'))\,\pi_\Theta(\phiti_\te(p))\,,
\end{align}
and the commutation relation in a fixed fiber are unchanged.
This is even stronger than the wedge-local structure mentioned in the Introduction. Whereas in the vacuum representation, $\phi_\te(f)$ and $\phi_{-\te}(g)$ commute if the supports of $f$ and $g$ lie in a particular spacelike position depending on $\te$, in the thermal representation $\pi_\Theta$ we have commutation of $\pi_\Theta(\phiti_\te(f))$ and $\pi_\Theta(\phiti_{-\te}(g))$ for {\em arbitrary} supports of $f,g$.

Thus the represented algebra $\pi_\Theta(\Pol_\Theta)$ appears ``more commutative'' than its vacuum counterpart $\Pol_\Theta$, its fibers almost completely decouple at finite temperature. In the context of question Q3) from the Introduction, this decoupling is the mechanism by which the representation $\pi_\Theta$ becomes wedge-local despite violating the spectrum condition. This however comes at the price of a non-separable representation space, in contrast to the vacuum situation which is better behaved from this point of view. It provides further evidence to the effect that the requirement of a decent thermodynamic behavior can seriously restrict ``twist structures'', as also observed in supersymmetric theories  \cite{BuchholzOjima:1997_4,BuchholzLongo:1999}.

\small
\subsubsection*{Acknowledgements}
We gratefully acknowledge helpful discussions with D.~Buchholz, and the hospitality of the Erwin Schr\"odinger Institute in Vienna. Initial stages of this work have been supported by the Austrian Science Foundation FWF under the project P22929-N16. GL would also like to thank the organizers of the ``Quantum Mathematical Physics'' conference for the invitation to Regensburg, where this work was presented.

\end{document}